\documentclass[conference]{IEEEtran}
\IEEEoverridecommandlockouts
\usepackage{cite}

\usepackage{soul}
\usepackage{etoolbox}
\patchcmd{\thebibliography}{\section*{\refname}}{}{}{}

\makeatletter 
\renewcommand{\fnum@figure}{\footnotesize \textbf{FIG.~\thefigure}}
\makeatother 
\usepackage{amsmath,amssymb,amsfonts}
\usepackage{algorithmic}
\usepackage{graphicx}
\usepackage{textcomp}
\usepackage{xcolor}
\def\BibTeX{{\rm B\kern-.05em{\sc i\kern-.025em b}\kern-.08em
    T\kern-.1667em\lower.7ex\hbox{E}\kern-.125emX}}

\makeatletter
\def\bstctlcite{\@ifnextchar[{\@bstctlcite}{\@bstctlcite[@auxout]}}
\def\@bstctlcite[#1]#2{\@bsphack
  \@for\@citeb:=#2\do{%
    \edef\@citeb{\expandafter\@firstofone\@citeb}%
    \if@filesw\immediate\write\csname #1\endcsname{\string\citation{\@citeb}}\fi}%
  \@esphack}
\makeatother 

\begin{document}
\bstctlcite{IEEEexample:BSTcontrol}

\title{Physics-inspired Ising Computing with Ring Oscillator Activated p-bits \vspace{-10pt}} 
\author{\IEEEauthorblockN{$\textbf{\ddag}$ Navid Anjum Aadit${^{*}}$, $\textbf{\ddag}$ Andrea Grimaldi${^\dagger}$, Giovanni Finocchio${^{\dagger,\circ}}$ \\and Kerem Y. Camsari${^{*,\natural}}$}
\IEEEauthorblockA{
${^{\dagger}}$Department of Mathematical and Computer Sciences, Physical Sciences and Earth Sciences, \\University of Messina, Messina, Italy\\
${^{*}}$Department of Electrical and Computer Engineering, University of California, Santa Barbara,\\ Santa Barbara, CA, 93106, USA}
$\textsection$\textbf{$\textbf{\ddag}$ \textit{Equally contributing authors}}}

\maketitle

 \begingroup\renewcommand\thefootnote{\textbf{\textsection}}
\noindent\footnotetext{Corresponding authors:  $^\circ$gfinocchio@unime.it,  $^\natural$camsari@ece.ucsb.edu}
\endgroup
\vspace{-10pt} 

\begin{abstract}
The nearing end of Moore's Law has been driving the development of domain-specific hardware tailored to solve a special set of problems. Along these lines, probabilistic computing with inherently stochastic building blocks (p-bits) have shown significant promise, particularly in the context of hard optimization and statistical sampling problems. p-bits have been proposed and demonstrated in different hardware substrates ranging from small-scale stochastic magnetic tunnel junctions (sMTJs) in asynchronous architectures to large-scale CMOS in synchronous architectures. Here, we design and implement a truly asynchronous and  medium-scale p-computer (with  $\approx$ 800 p-bits) that closely emulates the asynchronous dynamics of sMTJs in Field Programmable Gate Arrays (FPGAs). Using hard instances of the  planted Ising glass problem on the Chimera lattice, we evaluate the performance of the asynchronous architecture against an ideal, synchronous design that performs parallelized (chromatic) exact Gibbs sampling. We find that despite the lack of any careful synchronization, the asynchronous design achieves parallelism with comparable algorithmic scaling in the ideal, carefully tuned and parallelized synchronous design. Our results highlight the promise of massively scaled p-computers with millions of free-running p-bits made out of nanoscale building blocks such as stochastic magnetic tunnel junctions.
\end{abstract}

\begin{IEEEkeywords}
p-bits, combinatorial optimization, planted Ising, Chimera lattice, asynchronous computing, massive parallelism, magnetic tunnel junctions
\end{IEEEkeywords}

\section{Introduction}

With the nearing end of Moore's Law, domain-specific hardware and architectures are growing rapidly.
The notion of performing \textit{some tasks} more efficiently (area, speed and/or energy) rather than improving performance for \textit{general purpose} computing has led to the proliferation of special-purpose accelerators. With their widespread use, hard optimization problems have been a primary target of this approach and a variety of different  domain-specific hardware architectures have emerged (see, Ref.~\cite{mohseni2022ising} for a general and recent review). 

As an example of this growing trend, probabilistic bits or p-bits were introduced \cite{camsari2017stochasticL} as a building block which can accelerate a broad family of algorithms including Monte Carlo, Markov Chain Monte Carlo \cite{kaiser2021benchmarking}, Quantum Monte Carlo, statistical sampling for Bayesian inference and Boltzmann machine learning \cite{kaiser2022hardware} methods. p-bits have been shown to be compatible with powerful optimization techniques such as parallel tempering \cite{grimaldi2022spintronics} with competitive performance relative to all other Ising machines (classical and quantum) in select problems such as integer factorization and Boolean satisfiability \cite{aadit2021massively}. Their combination with sophisticated algorithms  \cite{mohseni2021nonequilibrium} could yield further advantages.  

A natural advantage of the p-bit model is its native mapping to the Ising model and to the natural generalization of Ising Models. This ensures that coupled p-bits can systematically probe the exact Boltzmann distribution through Gibbs or Metropolis sampling without any approximations or reductions, often necessary in alternative, non-bistable abstractions of the Ising spin.

One particularly promising small-scale demonstration of p-bits in an \textit{asynchronously} operating mode was performed in Ref.~\cite{borders2019integer}. Combined with key breakthrough experiments demonstrating nanosecond fluctuations in suitably designed low barrier magnetic tunnel junctions (MTJ) \cite{hayakawa2021nanosecond,safranski2021demonstration}, these results suggest the intriguing possibility of designing $>$ million bit probabilistic computers \cite{sutton2020autonomous} in light of the remarkable advances in the magnetic memory chip industry reaching gigabit densities \cite{finocchio2021promise,bhatti2017spintronics}. Even though large scale p-bit emulators have been designed and tested in FPGAs or ASICs, 
\cite{pervaiz2018weighted,smithson2019efficient,sutton2020autonomous,aadit2021massively,kaiser2021benchmarking}, virtually all of these implementations have been on \textit{synchronous} hardware where a global clock controlled the information flow. 

\begin{figure*}[t!]
\begin{center}
    \vspace{-10pt} 
	\includegraphics[width=0.80\linewidth]{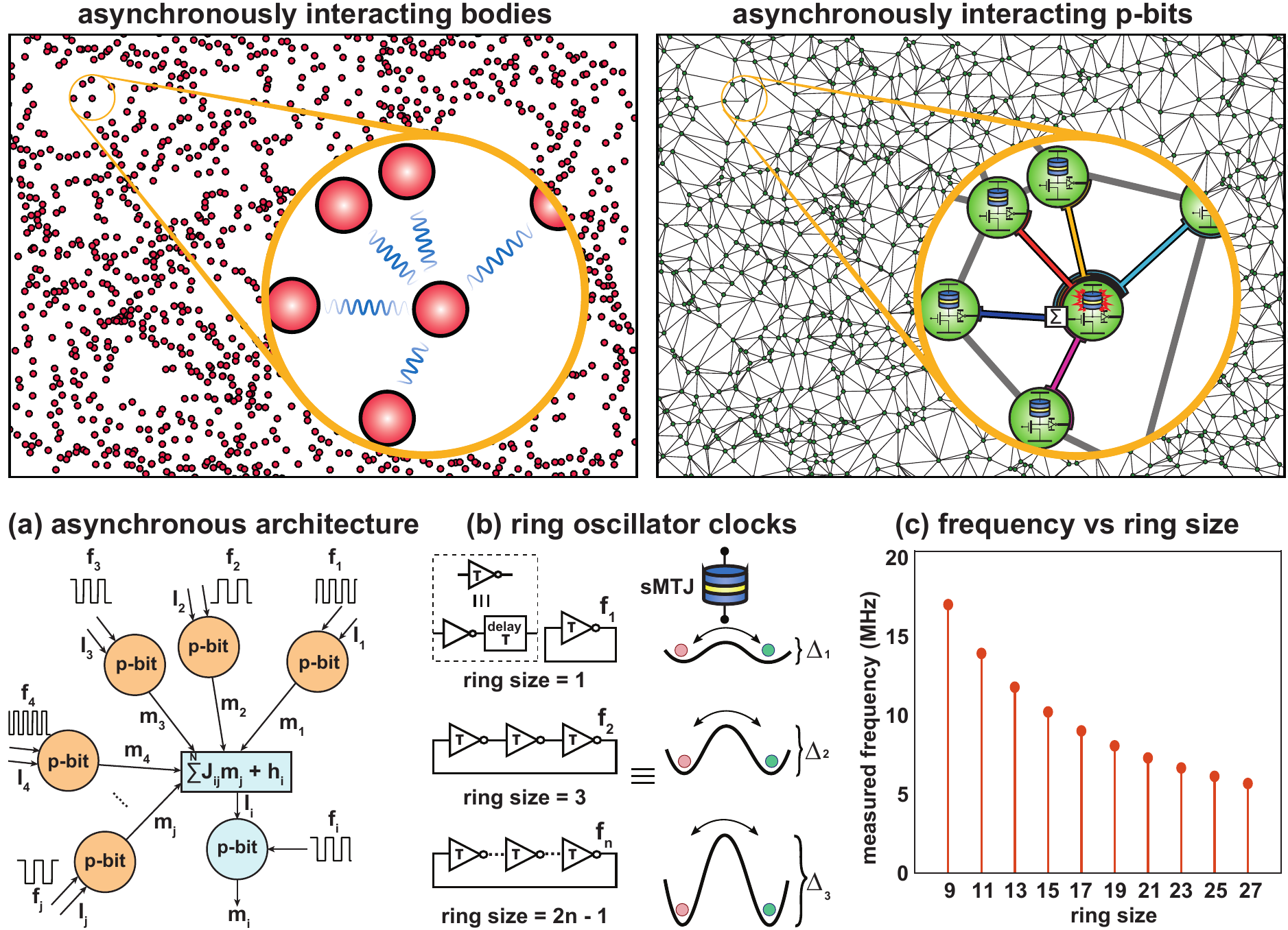}
	\caption{Upper panel: Physics-inspired analogy between asynchronously interacting bodies  and p-bits: both systems are asynchronous,  local (sparse connectivity) and massively parallel.  (a) Asynchronous computer architecture: The local field (Eq.~\eqref{eq:pbit}) for each p-bit is computed combinationally. Each p-bit has a different clock with asynchronous activation
    (b) Asynchronous ring oscillator-based clocks (ROSC) where we draw an analogy between stochastic Magnetic Tunnel Junction (sMTJ)-based p-bits and ROSCs.  Large rings correspond to large energy-barrier sMTJs with higher retention times. (c) Measured frequency range for ROSCs with different ring sizes in the FPGA implementation.}
	\label{fig:fig1}
\end{center}
\vspace{-7mm}
\end{figure*}

In this paper, we make a first attempt in designing and building a physics-inspired, \textit{truly asynchronous} architecture, closely emulating the dynamics of interacting nanodevice-based p-bits. This physics-inspired architecture bears similarities  to locally interacting (sparsely connected) and asynchronous bodies with probabilistic dynamics (FIG.~\ref{fig:fig1}, upper panel). We achieve the design by an unconventional use of FPGAs where individual p-bits are activated by decoupled ring oscillators and can have overlapping and out-of-phase clocks with different frequencies. Considering how variations may influence individual p-bit behavior in magnetic tunnel junction based designs \cite{rahman2021variability} the behavior of asynchronous p-computers with built-in variations is worth investigating. 

To compare the performance of the truly asynchronous p-computer in the FPGA, we choose the planted Ising model where a hard optimization problem is generated with a planted solution \cite{hen2015probing,albash2018demonstration}, allowing a reliable evaluation of the asynchronous design with respect to exact Gibbs sampling.


\section{Physics-inspired Architecture}
\label{sec:ROSC}
The main equations of the p-bit model (FIG.~\ref{fig:fig1}a) involves stochastic activation and a local field (synapse) calculation, given by:
\begin{equation}
    m_i = \mathrm{sgn}(\mathrm{tanh}(\beta I_i) - r_U) \quad 
    I_i = \sum J_{ij} m_j + h_i  
    \label{eq:pbit} 
\end{equation}
where $m_i$ represents the bipolar p-bit state ($\pm 1$), $r_U$ is a uniform random number between $(-1,+1)$ and $[J],\{h\}$ are the weights and biases for a given problem and $\beta$ is the inverse temperature. 

Standard Gibbs sampling iterates Eq.~\eqref{eq:pbit} to reach the Boltzmann distribution defined by the weights and typically involves a  \textit{serialized} update procedure with nested \texttt{for} loops.  One way to avoid this serial \texttt{for} loop is to perform block updates between unconnected p-bits. This approach when applied in software is named ``chromatic sampling'' \cite{gonzalez2011parallel} and a low-level hardware realization of it was recently reported in Ref.~\cite{aadit2021massively}. However, this design also involves carefully designed and equally phase shifted synchronous clocks so that multiple blocks do not update simultaneously. 

In this work, inspired by truly asynchronous small-scale implementations of p-computers with nanodevices (based on stochastic MTJs \cite{borders2019integer,kaiser2022hardware}),  we implemented a physics-inspired, truly asynchronous Ising Computer where different p-bits receive clocks with different frequencies with random phases. In contrast to synchronous designs, no careful engineering between the clocks of asynchronous p-bits were made. Moreover, unavoidable variations of sMTJs in highly scaled p-computers with nanodevices would make such engineering extremely difficult if not impossible. We found that despite the deliberate randomization of p-bit clocks and unavoidable collisions breaking exact Gibbs sampling, the physics-inspired design exhibited massive parallelism observed in carefully tuned synchronous designs, not observed in standard CPU-based Gibbs sampling (FIG.~\ref{fig:fig3}). 

\begin{figure*}[t!]
    \centering
    \includegraphics[width=0.85685\linewidth]{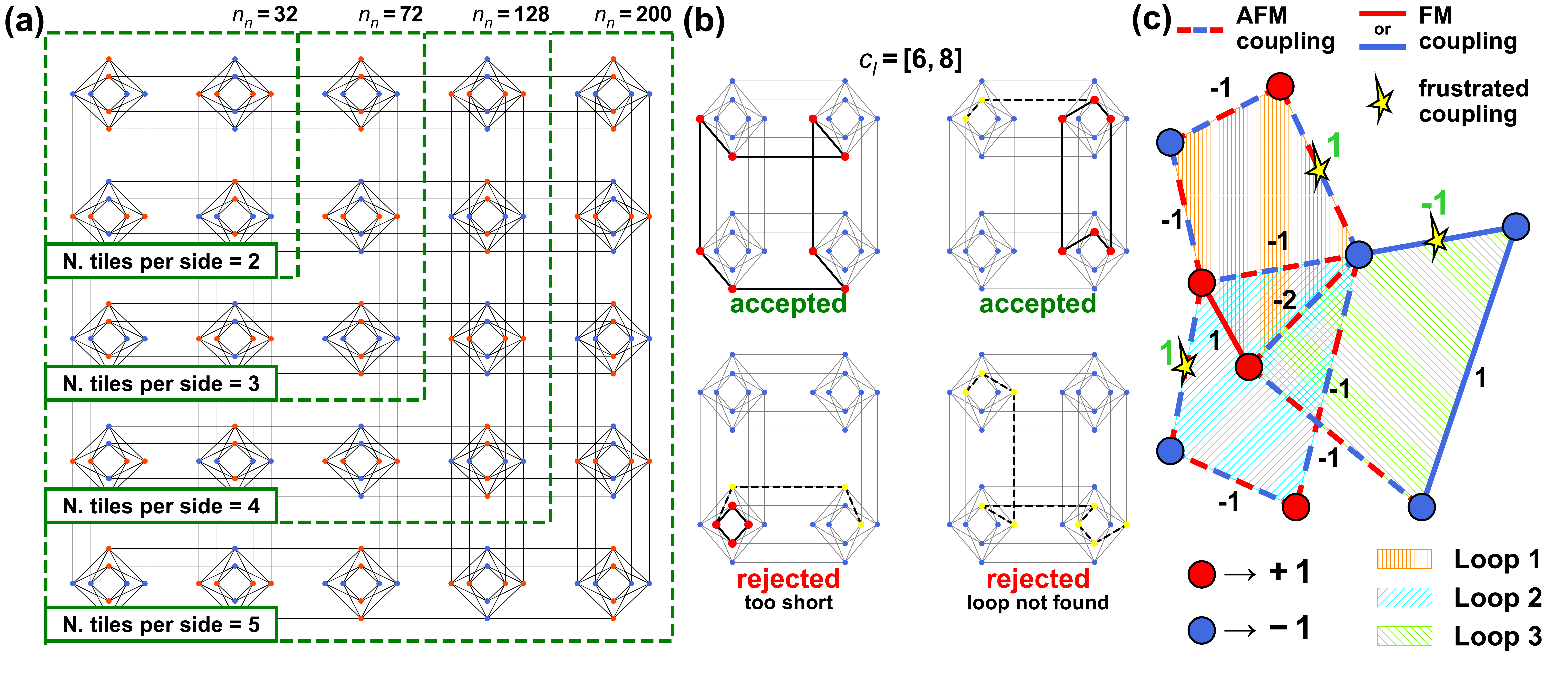}
    \vspace{-3mm}
    \caption{Planted Ising problems. (a) Chimera lattice with different sizes, depending on number of tiles. Different tiles were chosen in the same hardware to change problem sizes, similar to Ref.\cite{hen2015probing} but all problems used a fixed 800-spin Chimera in our hardware.  
    (b) Loop/close generation process shown for  $c_l=[6,8]$. (c) Example graph with three loops (the colored hatchings) and a planted solution (the colors of the nodes). For each loop, the weights of the couplings between neighbors are assigned to be either ferromagnetic (FM) interactions or antiferromagnetic (AFM), based on the plant. One randomly selected coupling in each loop is flipped (spark symbol).}
    \label{fig:chimera}
    \vspace{-15pt} 

\end{figure*}

\noindent \textbf{Ring Oscillator Generation:} A ROSC clock consists of an odd number of looped NOT gates. In our FPGA (Xilinx, VCU118), we attach controllable delays to our inverters to make logical delays comparable to wire delays (FIG.~\ref{fig:fig1}b). We designed the delay unit as a flip flop with a very fast master  clock ($300 $ MHz) compared to the ROSC frequencies that essentially acts as a combinational delay unit.  In this way, we were able to obtain highly regular ROSC clocks as a function of ring sizes whose frequencies were measured by specially designed counters (FIG.~\ref{fig:fig1}c).  In our experiments, we used 10 ROSCs to drive 800 p-bits in a Chimera lattice. Each p-bit has a pseudorandom number generator, which is a 32-bit Linear Feedback Shift Register (LFSR). The ROSCs activate the LFSRs of the p-bits randomly based on the frequencies. In this work, we have distributed the clocks evenly among the p-bits between 5 and 17 MHz. However, different distributions for the clocks, e.g., Gaussian, could be used. Since the Chimera graph is bi-partite, we did not assign the same clock to two p-bits that are on the different partitions to avoid systematic parallel updates between connected p-bits. Future work will consider \textit{dynamic} clocking schemes where each p-bit can have a different ``retention time'' much like MTJ-based p-bits.  \vspace{-6pt}

\section{Planted Ising Model}
An important class of hard optimization problems are those with ``planted'' ground states that allow effective evaluation of performance. We construct frustrated spin glasses with planted solutions \cite{hen2015probing} on a 800-spin Chimera graph  where we changed the number of tiles for different problem sizes (FIG.~\ref{fig:chimera}a). A Hamiltonian generated by this process is the sum of several local frustrated Hamiltonians which we will call ``clauses''. A planted solution will be used to define these clauses so that it will be the ground energy of the final Hamiltonian.
Every instance is characterized by two parameters: the clause density $\alpha$, defined as $n_c / n_n$, where $n_c$ is the number of clauses and $n_n$ is the number of nodes of the graph, and the length of possible loops that form the clauses, $c_l=[l_{min}, l_{max}]$, where $l_{min/max}$ is the min/max loop length, respectively. In this work, we chose $\alpha=0.4$ and $c_l=[4,8]$ for all our instances used in this paper, that run on the same 800-spin Chimera lattice in our hardware. \vspace{3pt}

\noindent \textbf{Clause generation:}
A total of $n_c=\alpha n_n$ clauses is generated. Each clause is an ordered sequence of nodes that creates a loop of acceptable length in the graph. To obtain one, following Ref.~\cite{hen2015probing}, we pick a random node and start a non-backtracking random walk of at most $l_{max}$ steps. If the walker lands on an already visited node, it means that a loop was formed and the node can be considered its initial point.
If the length of the loop is $>l_{min}$ the clause is accepted, if it is not or if the maximum number of steps is reached without closing the loop, the process is repeated. FIG.~\ref{fig:chimera}b shows a few examples of this process. 
A planted solution $s$ is generated by creating a random array of $-1$s and $+1$s of length $n_n$. A clause can be defined as $c_m=\{n_1, n_2, ..., n_k, n_{k+1}\}$, with $k \in [l_{min}, l_{max}]$, where $n_{k+1}=n_1$, representing the closing of the loop. 
Now, $\forall i \in [1, k]: i\neq j$ we increase $J_{n_i, n_{i+1}}$ by $s_i s_{i+1}$, while for $j \in [1, k]$, picked at random, we increase $J_{n_j, n_{j+1}}$ by $-s_j s_{j+1}$. This last step serves to create a frustrated loop. 
Once this is done for all clauses, the final $J$ is calculated by summing $J$ and $J^{T}$ and by normalizing so that all $J$ lie between $[-1, +1]$.
 \vspace{-15pt}

\section{Performance Comparison}
We follow the time-to-solution formulation\cite{hen2015probing, albash2018demonstration}
 to measure performance of the physics-inspired asynchronous architecture. 
\begin{equation}\label{EQ:TTS_Definition}
    \mathrm{TTS}(\tau, p_R) = \tau N_{r}(\tau, p_R)
\end{equation}
where $N_r$ is the expected number of repetitions we need to perform an annealing schedule of time $\tau$ to  the energy ground state at least once with probability $p_R$. $N_r$ is defined as:
\begin{equation}\label{EQ:N_R_Definition}
    N_r(\tau, p_R) = \frac{\ln{(1-p_R)}}{\ln{(1-p_S(\tau))}}
\end{equation}

\begin{figure}[!t]
\begin{center}
	\includegraphics[width=0.95\columnwidth]{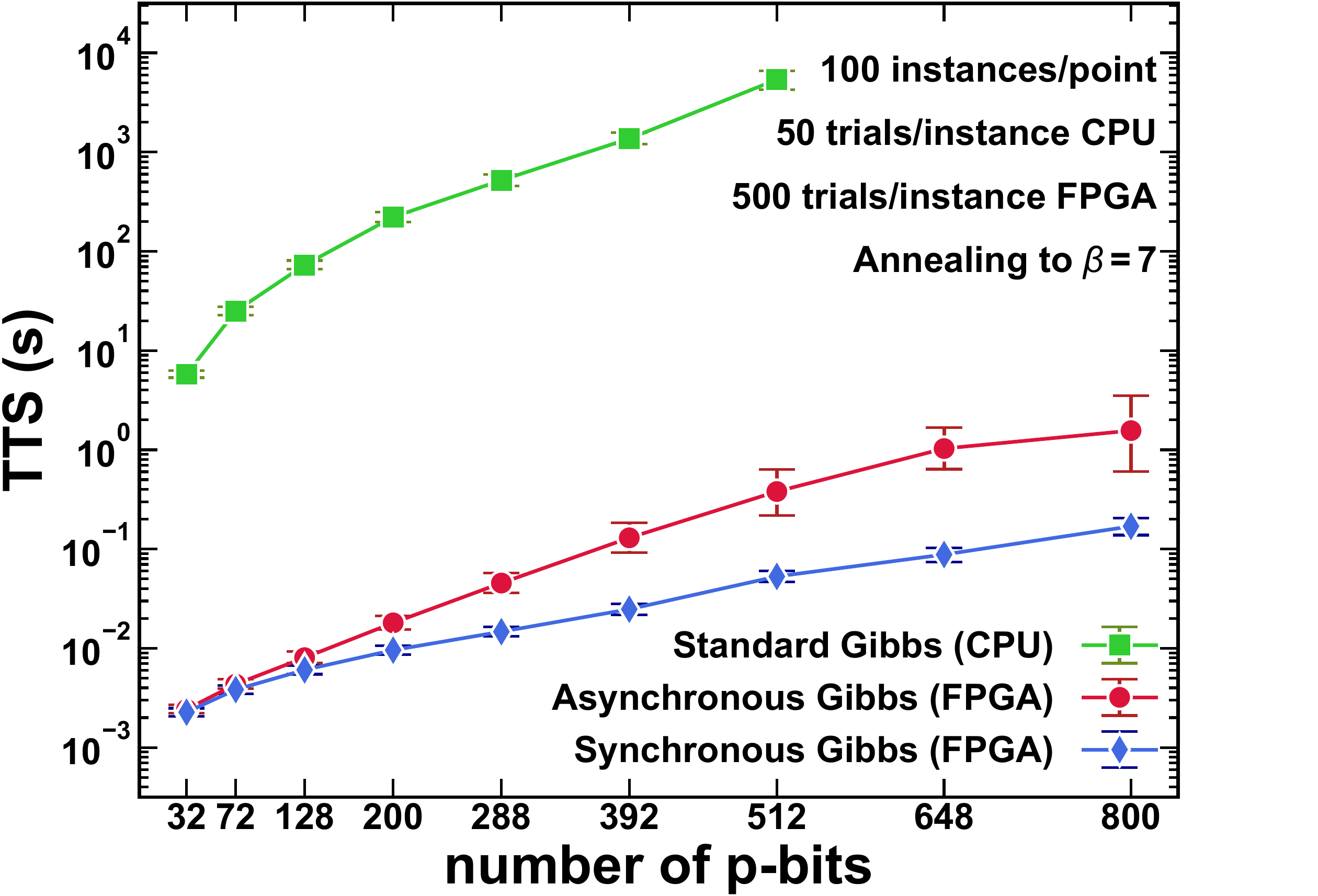}
   	\vspace{-1mm}
	\caption{Performance comparison of synchronous (graph colored-tuned), asynchronous (physics-inspired, not colored, not tuned) and standard Gibbs (CPU) samplers on the planted Ising problem. }
	\label{fig:fig3}
\end{center}
\vspace{-7mm}
\end{figure}

\noindent where $p_S$ is the probability of success in finding the ground state in one annealing process of length $\tau$. 

To evaluate the performance of our asynchronous architecture, we compared it to serialized Gibbs sampling on CPU ($2.6$ GHz) and to synchronous colored Gibbs sampling on FPGA. We investigated the scaling difficulty of planted Ising instances with fixed $c_l$ and $\alpha$ across several Chimera graphs increasing in size by changing the number of tiles used in a Chimera, as illustrated in FIG.~\ref{fig:chimera}.  
For each set of tiles,  we generated 100 planted Ising instances, performing simulated annealing trials to estimate the success probability, $p_S (\tau)$, for each instance (FIG.~\ref{fig:fig3}). On FPGA, we performed 500 trials per instance while on CPU we only performed 50 trials because of the exceedingly long run-times.  

After estimating a $p_S$ for every instance, the average $\mathrm{TTS}$ for a point is obtained by simply calculating the average of the $p_S$ for instances of that size and applying Eq.~\eqref{EQ:TTS_Definition} with the appropriate annealing time $\tau$. The reference probability $p_R$ was set to $99\%$. The error bars are obtained through bootstrapping with 95\% confidence intervals and $10^4$ samples.

We present the $\mathrm{TTS}$ for solving 100 instances of the planted Ising problems of 9 different sizes in FIG.~\ref{fig:fig3}. The standard Gibbs (CPU) was implemented in Python using optimized libraries for matrix calculations. The final two points were not computed because of time limitations. We solved the exact same instances on the FPGA programmed with the asynchronous ROSC activated 800 p-bits with a fixed point representation using 10-bits. As a reference, the \textit{synchronous} solver performing chromatic Gibbs sampling solves the same instances on the same FPGA where careful phase shifting ensures  no simultaneous or incorrect updates (where $I_i$ calculation is not complete) occur between neighboring p-bits (as in Ref.~\cite{aadit2021massively}). On the other hand, the asynchronous solver is expected to take samples with both of those errors when p-bit clock edges happen to be closely separated. Our experiment investigates the usefulness of such  samples. 

We defined a common linear simulated annealing schedule for all the architectures with $\beta = 0.5$ to $7.0$ with a step of $0.5$ where, at each $\beta$, a total of $937$ sweeps (attempted flips of all p-bits) are executed. In the FPGA the annealing time was fixed to $\tau = 1.4$ ms for each trial.
To obtain a configuration comparable to the asynchronous architecture, the synchronous architecture was set up with two stable and oppositely phase shifted clocks with the average frequency (9.375 MHz) of the 10 ROSC clocks. 
We believe this arrangement made the two designs equivalent beyond the \textit{asynchronous and inexact dynamics} of the ROSC since both designs approximately take the same amount of samples within the fixed annealing time $\tau$.
The key result we obtained is shown in FIG.~\ref{fig:fig3}. We observe a clear scaling difference between the CPU implementation of standard serialized Gibbs sampling and the massively parallel FPGA implementations which (ideally) obtain a scaling factor of $\approx N$ in their flips/second due to their massively parallel architecture. Both solvers  provide a roughly 5-orders of magnitude prefactor improvement over the CPU. Intriguingly, the scaling of the synchronous and asynchronous FPGA remain similar, despite the possibility of many collisions (parallel or incorrect updates) in the asynchronous design. Indeed, the carefully tuned synchronous design performs strictly better than the asynchronous one in all instances. Nevertheless, it is encouraging to observe that the asynchronous design without any carefully engineered clocks or tuning performs nearly as well, leading to the promising possibility of truly asynchronous, million bit p-computers with stochastic MTJs or other nanodevices.

\vspace{-3pt} 
\section*{Acknowledgment}
\vspace{-3pt} 

\footnotesize K.Y.C. and N.A.A. acknowledge support through National Science Foundation (CCF 2106260) and K.Y.C. through the Samsung GRO program. A.G. and G.F. were supported under the project PRIN
2020LWPKH7 funded by the Italian Ministry of University
and Research and by the PETASPIN Association (www.petaspin.com).  \vspace{5pt} 
\normalsize


\end{document}